\documentclass[manuscript,screen]{acmart}
\usepackage{color}
\usepackage{multirow}
\usepackage{makecell}
\usepackage{subfig}
\usepackage{colortbl}
\usepackage{xcolor}
\usepackage{pgfplots}
\usepackage{pgfplotstable}
\usepackage{appendix}
\usepackage{graphicx}
\pgfplotsset{compat=1.18}
\pgfplotsset{compat=1.18}
\usepackage{caption}
\usepackage{enumitem} 
\usepackage{tabularx}
\usepackage{array}
\usepackage{graphicx}
\usepackage{booktabs}

\AtBeginDocument{%
  \captionsetup[figure]{format=plain,justification=raggedright,
                        singlelinecheck=false,width=\textwidth}%
}

\usepackage{tabularx}   

\newcolumntype{L}{>{\RaggedRight\arraybackslash}X}

\newcommand{\getcolor}[3]{%
    \pgfmathsetmacro{\colorValue}{100.0*(#1-#2)/(#3-#2)}%
    \edef\temp{\noexpand\cellcolor{blue!\colorValue}\textcolor{black}{#1}}%
    \temp
}

\author{Lyumanshan Ye}
\affiliation{%
  \institution{Shanghai Jiao Tong University}
  \city{Shanghai}
  \country{China}
}
\email{yelvmanshan@gmail.com}

\author{Jiandong Jiang}
\affiliation{%
  \institution{Shanghai Jiao Tong University}
  \city{Shanghai}
  \country{China}
}

\author{Yuhan Liu}
\affiliation{%
  \institution{University of Copenhagen}
  \city{Copenhagen}
  \country{Denmark}
}
\email{yuhan.liu.connect@gmail.com}

\author{Yihan Ran}
\affiliation{%
  \institution{Shanghai Jiao Tong University}
  \city{Shanghai}
  \country{China}
}

\author{Yufan Zhou}
\affiliation{%
  \institution{KU Leuven}
  \city{Leuven}
  \country{Belgium}
}

\author{Zhao Wang}
\affiliation{%
  \institution{Zhejiang University}
  \city{Hangzhou}
  \country{China}
}

\author{Yipeng Yu}
\affiliation{%
  \institution{Taotian, Alibaba}
  \city{Shanghai}
  \country{China}
}

\author{Pengfei Liu}
\affiliation{%
  \institution{Shanghai Jiao Tong University}
  \city{Shanghai}
  \country{China}
}
\email{pengfei@sjtu.edu.cn}

\author{Danni Chang}
\affiliation{%
  \institution{Shanghai Jiao Tong University}
  \city{Shanghai}
  \country{China}
}
\email{dchang1@sjtu.edu.cn}

\author{Yucheng Jin}
\affiliation{%
  \institution{Duke Kunshan University}
  \city{Suzhou}
  \country{China}
}
\email{yj232@duke.edu}

\begin{document}

\title[\textit{Colin}]{\textit{Colin}: A Multimodal Human-AI Co-Creation Storytelling System To Support Children's Narrative Skills}

\renewcommand{\shortauthors}{Ye et al.}

\begin{abstract}
Children develop narrative skills by understanding and actively building connections between elements, image-text matching, and consequences. However, it is challenging for children to clearly grasp these multi-level links only through explanations of text or the facilitator's speech. To address this, we developed \textit{Colin}, an interactive storytelling tool that supports children's multi-level narrative skills through both voice and visual modalities. In the generation stage, \textit{Colin} supports the facilitator to define and review the generated text and image content freely. In the understanding stage, a question-feedback model helps children understand multi-level connections while co-creating stories with \textit{Colin}. In the building phase, \textit{Colin} actively encourages children to create connections between elements through drawing and speaking. A user study with 20 participants evaluated \textit{Colin} by measuring children's engagement, understanding of cause-and-effect relationships, and the quality of their new story creations. Our results demonstrated that \textit{Colin} significantly enhances the development of children's narrative skills across multiple levels.

\end{abstract}

\begin{teaserfigure}
\centering
\includegraphics[width=1\textwidth]{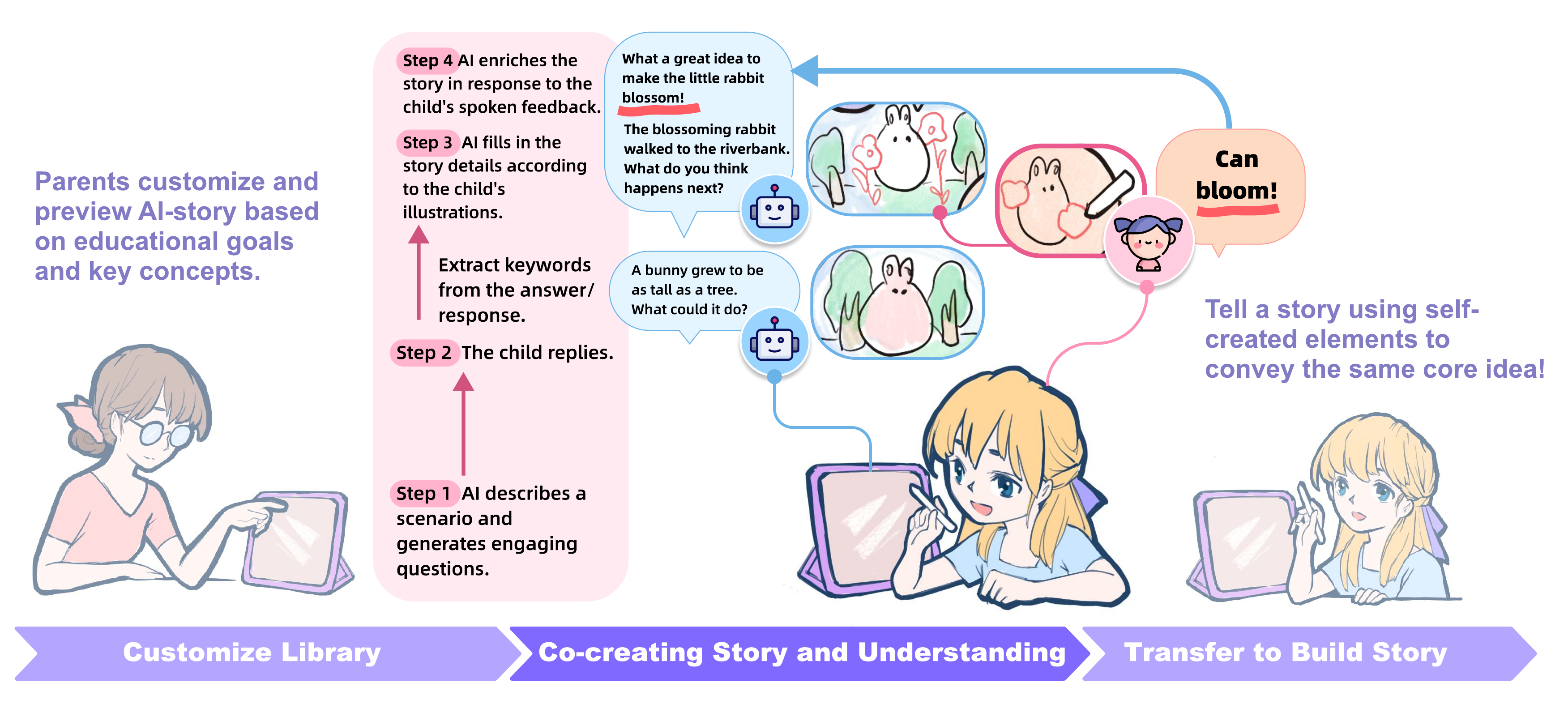}
\caption{Workflow of the LLM-empowered multimodal storytelling system. Parents set educational goals and preview stories. Children co-create with AI through drawing and dialogue, and finally retell a story using their own elements to express the same core idea.}
\label{fig:colin}
\end{teaserfigure}

\maketitle



\section{Introduction}

Storytelling plays a foundational role in children’s cognitive, linguistic, and socio-emotional development. Through constructing and interpreting stories, children learn to organize events chronologically, express causality, and derive meaning—skills essential for reasoning about intentions, emotions, and social relationships~\cite{glaser2009narrative, cleeve2024reflecting}. Developing narrative competence in early childhood thus lays the groundwork for imagination, empathy, and higher-order thinking.

Interactive storytelling, a child-centered pedagogical approach, integrates narrative with interaction and has become a core method in early education~\cite{cremin2017storytelling}. Facilitators engage children through dialogic exchanges, posing story-related questions and scaffolding responses using strategies such as repetition, elaboration, and modeling~\cite{farini2023my}. These interactions enable children to co-construct narratives, articulate ideas, and visualize abstract concepts through drawing~\cite{wu2012pedagogical}. However, because this approach requires individualized attention, large teacher–student ratios often hinder sustained practice in classrooms, prompting many parents to adopt storytelling-based learning at home. Parents extend story contexts and link them to real-life experiences, which has been shown to significantly enhance children’s narrative development~\cite{schmidgall2019learners}. Yet, sustaining interactive storytelling at home presents major challenges. Parents must improvise storylines in real time, balance guidance with autonomy, and visualize story scenes—tasks that demand creativity, contextual knowledge, and expressive skill. Repetition, though beneficial for learning, can also be time-consuming and tiring. Moreover, many parents lack the pedagogical strategies for responsive scaffolding—the ability to adapt to a child’s creative input while maintaining coherence~\cite{good2006carss}—and may struggle with domain knowledge or illustration skills~\cite{mishra2020genre, kendeou2020inferential}. These difficulties often limit parents’ capacity to maintain consistent storytelling practices, especially in digital or AI-mediated contexts~\cite{Pillinger2022, Nan2025, Sun2024}.

Recent advances in large language models (LLMs) and multimodal generative AI offer new opportunities to address these challenges. Unlike earlier systems that relied on predefined narrative structures~\cite{szilas2007computational, riedl2013interactive}, modern generative models can engage in contextually relevant, adaptive dialogue and produce creative multimodal content~\cite{pirjan2025artificial, zhang2024unveiling}. Such systems can augment parents’ pedagogical, linguistic, and artistic capacities, creating new opportunities for human–AI co-facilitation in children’s learning~\cite{teepe2017technology}. However, how LLMs can effectively support co-creative storytelling between parents and children—particularly in multimodal and educational contexts—remains underexplored.

To address this gap, we present \textit{Colin}, a multimodal AI-powered storytelling system that supports collaborative story co-creation between parents and children. \textit{Colin} leverages LLMs and generative models to scaffold narrative learning through dialogue and drawing, providing adaptive, context-aware feedback to enhance story comprehension and creativity.
Building on a formative study with parents and education experts, we developed \textit{Colin} and conducted an empirical study with 20 parent–child pairs to explore how the system supports narrative co-creation and learning.
Through the design and evaluation of \textit{Colin}, we seek to address the following research questions:
\begin{itemize}
    \item RQ1: How can LLMs be effectively and appropriately used to support facilitator-involved scaffolding and co-creation in children’s narrative development?
    \item RQ2: How can adaptive interaction mechanisms influence children’s learning outcomes and learning experiences?
    \item RQ3: How do multi-stakeholders (children and parents) perceive an LLM-empowered storytelling system in supporting children’s narrative development?
\end{itemize}
	
Our findings suggest that a multimodal interactive storytelling system can enhance children’s narrative outcomes and engagement, strengthen parents’ control over educational goals, and accommodate diverse storytelling needs. Our work makes three primary contributions:
\begin{enumerate}
    \item The design and development of \textbf{\textit{Colin}}, a multimodal storytelling co-creation system that employs question-scaffolding feedback to enhance story learning through conversational and sketch-based interaction.
    \item An instructional prompt-engineering framework for child–AI co-creative storytelling based on LLMs, evaluated through a comparative content generation task.
    \item Empirical insights into how LLM-assisted storytelling influences children’s learning experiences and how parents and children perceive AI-mediated co-creation in educational contexts.
\end{enumerate}



\section{Related works}
\subsection{Children's Narrative Skills and Technology Support}

The ability to narrate visual scenes is a core component of children’s narrative skills, requiring them to logically connect pictorial elements into coherent storylines. Visual narratives—such as picture books, comics, and other sequential image formats—present successive events with the intention of conveying a story~\cite{cohn2020visual,10.1145/3585088.3593896,10.1145/3526113.3545617,10.1145/3613904.3642492}. Contrary to the assumption that meaning in image sequences is self-evident~\cite{arbuckle2004language}, comprehension depends on the child’s ability to identify, relate, and infer links between visual elements to form an integrated narrative. This process, known as visual narrative comprehension\cite{cohn2020visual}, involves understanding continuity, causality, and activity across frames—skills that develop progressively throughout childhood\cite{haddad2015visual}. Narrative logic more broadly encompasses the organization of events through temporal sequencing, causal relations, and character interactions~\cite{pico2021interventions}. According to Stein and Glenn’s seminal story grammar framework~\cite{stein1979analysis}, coherent storytelling integrates seven key components—situation, trigger, internal response, plan, attempt, outcome, and reaction—whose coordination enables meaningful story construction and comprehension.

Building on these developmental foundations, HCI researchers have explored diverse technological approaches to scaffold children’s storytelling and narrative thinking. Systems such as StoryMat \cite{ryokai1999storymat} pioneered tangible and embodied storytelling, allowing children to record and replay narratives through interactive physical artifacts. Other work, such as Making Space for Voice \cite{cassell2001makingspace}, articulated broader design principles for voice-based support of children’s fantasy play and storytelling. More recently, researchers have begun integrating AI and language models into narrative interfaces. For example, STARie \cite{li2023starie} is a peer-like embodied conversational agent designed to co-construct stories with children using multimodal interaction. Another line of work explores AI-assisted narrative tools such as Mathemyths \cite{zhang2024mathemyths}, which co-creative storytelling with children incorporates target learning objectives (in that case, mathematical language). Systems like StoryBuddy \cite{zhang2022storybuddy} further illustrate human–AI collaborative storytelling with flexible caregiver involvement.

The above-mentioned systems highlight a growing movement in HCI toward designing co-creative storytelling environments that merge children’s expressive modalities—speech, drawing, and play—with adaptive computational scaffolds. This trajectory informs our work on \textit{Colin}, a multimodal human–AI co-creation system that supports children’s multi-level narrative development by connecting their verbal and visual expressions through generative AI feedback.

\subsection{AI Assistant Multi-model Storytelling Systems}
Digital storytelling, supported by visual narrative activities, is an effective educational tool that promotes critical literacy and creativity. By engaging learners in encoding and decoding visual elements, it deepens their understanding of visual communication and stimulates original thought ~\cite{williams2019attending,ware2005hybrid}. Increasingly, AI is integrated into digital storytelling systems to combine narration with visual depiction. For example, OSOS employs pervasive language modeling to extract priority words from a child's language environment and generate customized storybooks ~\cite{lee2024open}. Communics ~\cite{rutta2020collaborative} supports collaborative comic-based storytelling by enabling children to select graphical elements and text. Such systems empower children to influence story development through decision-making ~\cite{cavazza2002planning}. Similarly, Fiabot! groups story elements by genre, allowing children to import and edit diverse media to build narratives, effectively stimulating discussion on plot and characters ~\cite{10.1145/2593968.2593979}. However, many systems rely heavily on children's visual thinking (e.g., arranging all elements on a single page), which may limit narrative completeness in sequential contexts. As Strube argues, relevance and coherence are crucial for constructing comprehensible visual narratives ~\cite{strube2010telling}. Hubbard further shows that robot-based visual interactions, supported by open-ended questions and dialogue, can foster children's coherent expression and creativity ~\cite{hubbard2021child}.

Interactive storytelling has long been shown to improve children's language and narrative skills ~\cite{gallagher2011search,collins1999use,bonds2016best,leonard1990storytelling,anggryadi2014effectiveness,zuhriyah2017storytelling}. In HCI, prior work in NLP has focused on system performance in interactive storytelling ~\cite{10.1145/3491102.3517479,yao-etal-2022-ais,marzuki2016improving,xu2022fantastic}, with less attention to how children engage in the learning process. Recently, large language models (LLMs) have been introduced into children's storytelling. Automated planning techniques help generate coherent and believable stories ~\cite{simon2022tattletale,simon2024does,chen2023fairytalecqa,chen2023fairytalecqaintegratingcommonsenseknowledge}. Beyond text generation, LLMs are being explored for supporting cooperative activities ~\cite{PeerGPT,zha2024designing}, parent-child interaction ~\cite{lee2022interactive,10.1145/3628516.3655809,10.1145/3501712.3529734}, and children's social engagement ~\cite{seo2023chacha}. While parents express strong needs for interactive systems ~\cite{sun2024exploring}, tangible interface approaches remain underexplored.

\subsection{Children-AI Collaborative Story Generation}
Researchers have designed diverse story idea generation tools for children, including virtual reality avatars and voice interfaces ~\cite{barber2010generation}, drawing-based and tangible interfaces ~\cite{feishu-liang2021tangible,10.1145/3628516.3659386,10.1145/3586183.3606807}, and graphical toolkits ~\cite{shao2021storyicon,murray2017proposal}. These studies highlight the value of multimodal creative support, enabling children to express stories through multiple mediums. Yet, fewer works explore how participatory creation fosters children's understanding of narrative structure. Benton's co-creation game ~\cite{benton2014understanding} and Zarei's enactment-scaffolded authoring tools ~\cite{zarei2021towards} demonstrate narrative collaboration but overlook educational goals. Other interactive systems show benefits for math ~\cite{zhang2024mathemyths,ruan2020supporting}, coding ~\cite{rocha2023coding,chatscratch,10.1145/3544548.3580981,10.1145/3411764.3445039}, and narrative skills ~\cite{liang2015exploitation}.

AI-assisted storytelling systems have further expanded children's creative possibilities. For example, StoryDrawer converts children's oral narratives into real-time sketches ~\cite{zhang2022storydrawer}, transformer-based models decompose story generation tasks ~\cite{wang2019quick}, and AnimaChaotic generates 3D animations from textual input ~\cite{abdel2022animachaotic}. StoryIcon integrates conversational agents to support role-play ~\cite{shao2021storyicon}, while Narratron uses gesture recognition, pretrained language models, diffusion-based visuals, and speech synthesis to produce multimodal stories ~\cite{10.1145/3586182.3625120}. These systems combine visual and linguistic cues, enabling children to create through drawing, narration, and dialogue, while strengthening motivation and engagement.

Recently, large language models (LLMs) have been introduced to education ~\cite{kasneci2023chatgpt}, supporting content generation ~\cite{dijkstra2022reading,gabajiwala2022quiz}, adaptive learning ~\cite{sailer2023adaptive,10.1145/3613904.3642379}, and personalized feedback ~\cite{jia2021all,el2021effects}. They have also been applied to children's storytelling, generating narratives ~\cite{alabdulkarim2021automatic,guan2020knowledge,10.1145/3613905.3651118} and serving as collaborative partners in dialogue or story continuation ~\cite{peng2023storyfier,10.1145/3613905.3651118}. Unlike prior systems that emphasize narrative richness, \textit{Colin} introduces two innovations: (1) employing visual modalities to intuitively represent story elements, and (2) integrating design features that enhance children's visual comprehension and narrative association. By interlinking pictorial elements, \textit{Colin} encourages children to expand discrete ideas into coherent, connected narratives.

\section{Formative Study}

To ground our system design in real-world educational practices, we conducted a formative study to examine how storytelling is applied and valued across both classroom and home contexts. We first interviewed five education experts, who were recruited for their professional experience in designing and implementing storytelling-based pedagogies. These experts shared how storytelling is integrated into formal education, the cognitive and emotional outcomes it aims to foster, and their recommendations for incorporating AI technologies into narrative learning experiences. To complement these institutional insights, we also interviewed three parents, as they play a primary role in children’s early narrative development and frequently use storytelling as an informal educational and bonding practice. The parent interviews explored how storytelling is embedded in daily routines, the challenges they face in sustaining engagement and educational value, and their expectations for AI-assisted storytelling tools in home settings. Together, these perspectives informed our design goals and highlighted key opportunities for human–AI collaboration to support children’s storytelling development.

\subsection{Methods}
\subsubsection{Participants}
This study involved two groups of participants: education experts and parents. We first recruited five education experts to validate the pedagogical soundness and practical feasibility of story-based teaching methods. The experts had an average age of 42. Three had over ten years of full-time early childhood education experience, and two had more than three years of professional teaching experience.
To further understand the implementation of storytelling education in family settings, we subsequently recruited four parents, each with children aged between 3 and 8 years old. These parents regularly engaged in storytelling activities such as reading picture books, using story machines, or narrating stories directly. Parents were selected because they can better observe and articulate children's cognitive and emotional responses during storytelling, which are often difficult for young children to express clearly. These two participant groups provided a comprehensive perspective on the use of interactive storytelling as an educational method—experts offered theoretical and pedagogical insights, while parents provided practical feedback grounded in everyday family experiences.

\subsubsection{Procedures}
We conducted semi-structured interviews with both participant groups. The interviews with education experts aimed to investigate (1) the educational purposes of using storytelling, (2) the concrete methods they employ in classroom teaching, (3) the developmental benefits and educational effects of interactive storytelling, (4) challenges in implementation, and (5) their perspectives on AI-assisted educational tools. The interview guide included questions such as:  
\textit{``Can you describe a specific example of how you use storytelling as a teaching method in your classroom?”} and  
\textit{``What challenges have you encountered when preparing and delivering story-based lessons, and how do you address them?”}  
These questions encouraged experts to share their experiences and reflect on the cognitive and pedagogical processes involved.  

The interviews with parents followed a similar structure, adapted to the family context. Parents were asked about (1) their educational goals when telling stories to their children, (2) their typical methods of interactive storytelling and perceived benefits for children, (3) the difficulties they encountered in maintaining interaction and engagement, and (4) their views on AI-assisted storytelling systems. Example questions included:  
\textit{``When you tell stories to your child, what do you hope they can learn or experience from it?”} and  
\textit{``What difficulties do you encounter when trying to engage your child in interactive storytelling?”}  
These questions guided participants to discuss both their intentions and practical challenges in depth.  

All interviews lasted approximately 60 minutes and were conducted either online or in person. Participants were encouraged to describe concrete storytelling scenarios, interaction strategies, and reflections on children's responses. The semi-structured design ensured flexibility to explore emerging insights while maintaining consistency across interviews.  
A full list of interview questions used in both expert and parent interviews can be found in \textbf{Appendix A}

\subsection{Key Insights}
The insights converge on four essential Key Insights (KIs): (1) storytelling serves both everyday knowledge transfer and the cultivation of abstract reasoning; (2) facilitators often face difficulties in scaffolding children's imaginative but fragmented contributions into coherent narratives; (3) personalized, multimodal support is needed to sustain engagement and foster independent thinking; and (4) story-based question guidance deepens children's comprehension and narrative thinking. 

\textbf{KI1: Adult Facilitation Is Essential for Developing Children's Narrative Skills.}  
Both experts and parents emphasized that children rely on adult guidance to construct coherent stories. Without structured support, children often struggle to organize events or express causal relationships. Teachers and parents scaffold storytelling through prompts that help recall key events, explain character motivations, and expand fragmented ideas—support that is critical for maintaining narrative coherence and meaning.  

\textbf{KI2: Integrating Children’s Creative Ideas into Story Flow Is Challenging.}  
Participants noted that children’s spontaneous and imaginative responses often disrupt story structure. Facilitators must quickly reinterpret or incorporate these ideas to sustain the narrative while preserving educational goals. This real-time balancing of creativity and coherence demands strong improvisational and domain skills, posing challenges even for experienced adults.  

\textbf{KI3: Multi-level Narratives Require Multi-modal Support.}  
Experts observed that children’s comprehension of complex stories—those involving shifting timelines, morals, or multiple characters—depends heavily on visual and auditory cues. Parents similarly found that illustrations, gestures, and voice modulation help sustain attention and clarify meaning. Both groups called for multi-modal aids, such as interactive visuals and voice-based hints, to better support understanding.  

\textbf{KI4: Adaptive Question Guidance Deepens Comprehension but Burdens Adults.}  
Story-based questioning effectively enhances children’s comprehension and moral reasoning, yet requires adults to repeatedly rephrase and adjust questions to sustain engagement. This repetitive task is cognitively demanding and time-consuming. Participants emphasized the need for AI systems that can dynamically generate varied, context-relevant questions to deepen understanding while easing facilitator workload.  

\subsection{Design Goals}
From these Key Insights, we distilled 4 Design Goals (DGs) that guided our system design process:

\begin{itemize}
    \item \textbf{DG1: Support facilitator preview in scaffolding.} Address the key insight that children's narrative development requires adult guidance \textbf{(KI1)} by enabling educators or parents to review and adjust AI-generated scaffolds before presenting them to children. 
    
    \item \textbf{DG2: Support co-creation through question scaffolds.} Tackle the difficulty of integrating children's ideas in real time \textbf{(KI2)} by embedding structured, open-ended questions that invite children to actively shape narratives. 
    
    \item \textbf{DG3: Enable multimodal narrative construction.} Respond to the need for clarifying multilayered narratives \textbf{(KI3)} by combining visual and auditory cues to scaffold children's understanding of causal and temporal connections. 

    \item \textbf{DG4: Enhance comprehension through adaptive question guidance.} Address the challenge that sustaining children's engagement and reinforcing story understanding requires repetitive questioning \textbf{(KI4)} by enabling the system to dynamically generate and adapt story-based prompts across three progressively deepening interaction phases—\textbf{Co-creation}, \textbf{Understanding}, and \textbf{Transfer}—to guide children's reflection on core ideas while reducing adults' facilitation burden.
\end{itemize}

These objectives serve as the conceptual bridge between formative insights and the subsequent system design.


\section{\textit{Colin}}

\subsection{Key Design Features}


 \begin{figure}[t]
     \centering
     \includegraphics[width=0.9\linewidth]{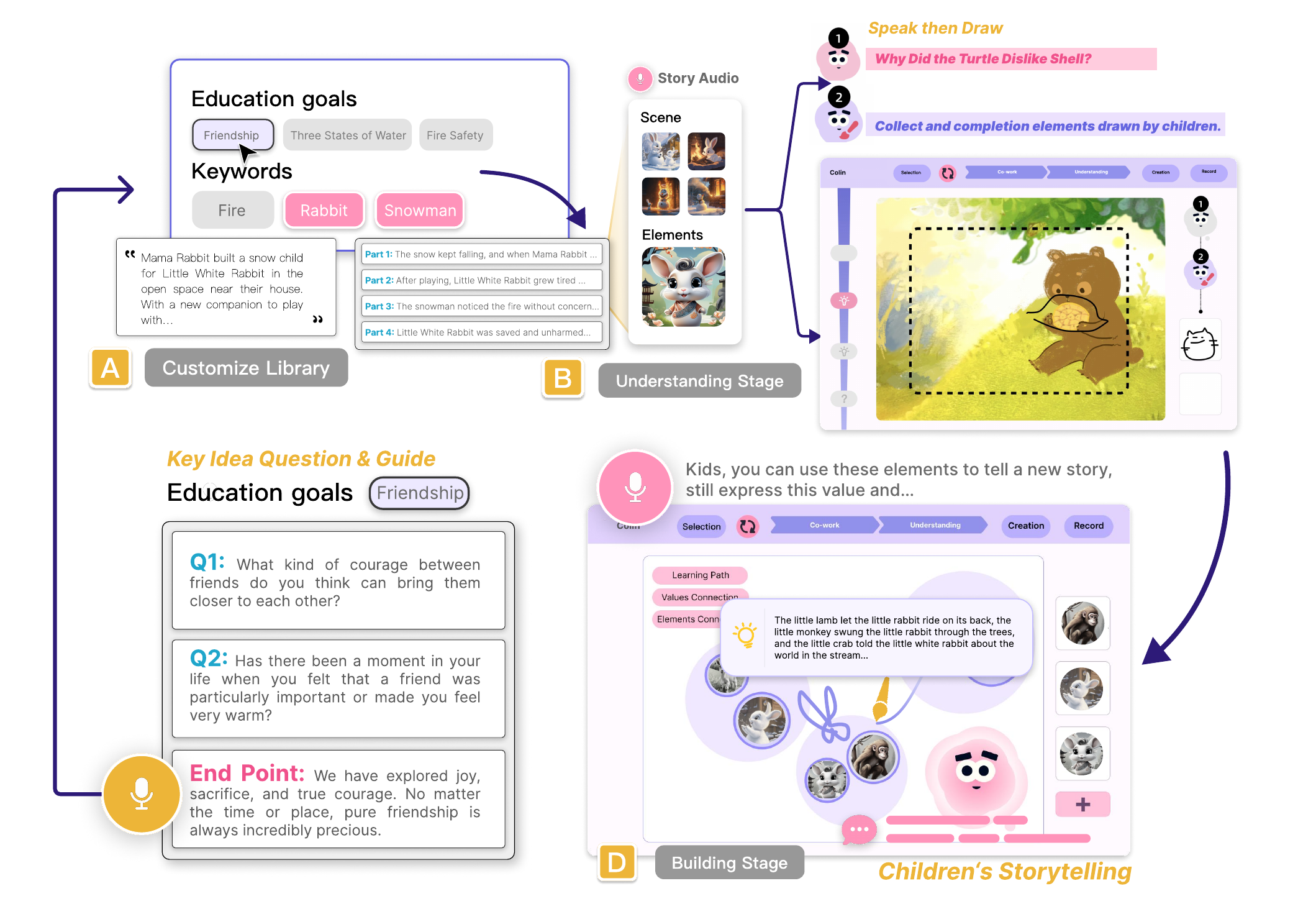}
     \caption{Overview of the AI-assisted storytelling co-creation process. \textbf{(A) Customize Library:} Parents set educational goals and keywords. \textbf{(B) Understanding Stage:} Children engage through speaking and drawing to interpret the story. \textbf{(C) Building Stage:} Children reuse created elements to build new stories expressing the same core value.}
     \label{fig:colin_process}
 \end{figure}

Guided by the identified design goals (DGs), we developed \textit{Colin}, a multimodal generative AI system designed to expand children’s narrative contributions, integrate adult facilitation, and promote deeper story understanding. \textit{Colin} is structured around four key features (\textbf{KFs}) that operationalize these design goals: (1) customizable and previewable story-framework generation (\textbf{KF1} addressing \textbf{DG1}); (2) LLM-driven conversational storyline expansion (\textbf{KF2} addressing \textbf{DG2}); (3) story-element creation through drawing (\textbf{KF3} addressing \textbf{DG3}); and (4) adaptive, progressively deep question guidance that links new elements to the evolving narrative (\textbf{KF4} addressing \textbf{DG4}).
Below, we elaborate on these key features and how they address the associated design goals. Figure~\ref{fig:colin_process} shows the key user interfaces and storytelling co-creation process of \textit{Colin}.

\textbf{KF1: Customizable and Previewable Story Framework Generation (DG1)}  
To support adult scaffolding in children’s narrative development \textbf{(KI1)}, \textit{Colin} provides a \textbf{Customizable Story Library} that enables facilitators to set educational goals and generate tailored story frameworks. Parents can input keywords or select from system-provided themes to create story scripts, each accompanied by four previewable scene illustrations (Figure~\ref{fig:colin_process}). They can edit text, regenerate visuals, and ensure pedagogical appropriateness before co-creation begins. Newly introduced creative elements from children’s interactions are stored in the same library, forming a “human-in-the-loop” design that aligns storytelling content with children’s evolving interests and learning progress.  

\textbf{KF2: LLM-Driven Conversational Storyline Expansion (DG2)}  
To address the challenge of integrating children’s spontaneous ideas in real time \textbf{(KI2)}, \textit{Colin} uses GPT-4 to guide children through two rounds of interactive storytelling. During the \textbf{Co-creating Story} phase, the system poses open-ended questions at predefined points, allowing children to record responses that are transcribed and incorporated into the evolving storyline (Figure~\ref{fig:promptflow}). Based on educator feedback, we limited the interactions to two questions to sustain attention while encouraging creativity. Rather than following a fixed plot, the LLM dynamically merges children’s imaginative inputs into the story, maintaining both coherence and engagement.  

\textbf{KF3: Story Element Creation through Drawing (DG3)}  
To enhance understanding through multimodal expression \textbf{(KI3)}, \textit{Colin} supports voice- and drawing-based interaction. During co-creation, children can open a sketchboard to visualize story elements mentioned in their narration. The “speak–draw–complete” mechanism leverages a \textit{sketch-rnn} model to recognize nouns from speech, generate visual elements, and assist with partial drawings. This multimodal process bridges linguistic narration and visual representation, deepening comprehension and memory. All generated visuals and drawings are collected for later reassembly in the “building stage,” where children rearrange elements to extend storylines and reinforce narrative logic.  

\textbf{KF4: Adaptive Question Guidance and Explanation (DG4)}  
To reinforce comprehension and sustain engagement \textbf{(KI4)}, \textit{Colin} adopts an adaptive question-guidance framework across three stages: \textbf{Co-creation}, \textbf{Understanding}, and \textbf{Transfer}. The system generates recall and feedback prompts to help children reflect on story meaning and correct misconceptions. In the transfer stage, children apply learned concepts by creating new stories using previous visual elements, sometimes combined with randomly introduced items to encourage creative generalization. All created materials are stored in the \textbf{Customize Library}, allowing parents to review progress and adjust future experiences. This adaptive mechanism promotes comprehension, creativity, and cross-context learning.  

\subsection{System Design and Workflow}

\textit{Colin} integrates multimodal elements—including images, speech, and text—and supports voice and visual interactions specifically tailored for children. We employed GPT-4 for story text generation and adapted the interface following the OpenAI API standards. To maintain conversational consistency, we organized and stored interaction histories in accordance with OpenAI's prompt rule recommendations. For image generation, we integrated the ChatGLM API~\cite{ChatGLMurl}.

\begin{figure}[htbp]
    \centering
    \includegraphics[width=0.9\linewidth]{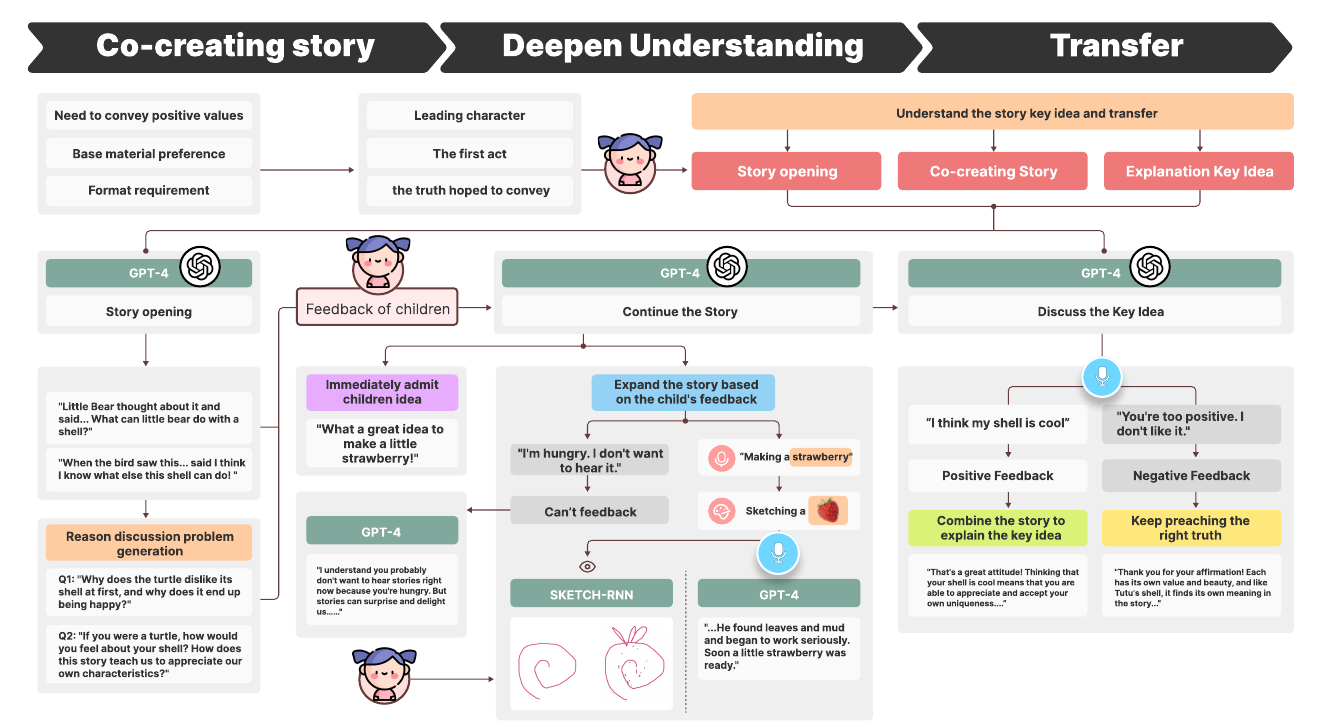}
    \caption{The overall workflow of \textit{Colin}, an AI-assisted storytelling system that leverages GPT-4 to co-create stories with children, guide understanding of key story values, and support knowledge transfer.}
    \label{fig:workflow}
\end{figure}

The overall workflow of \textit{Colin} is structured around five key interaction processes, as illustrated in Fig.~\ref{fig:workflow}. 
Before children use \textit{\textit{Colin}} for story co-creation, the \textbf{Customize Library} module allows parents to input desired educational goals and keywords to generate customized story scripts, or to select from system-provided goals and keywords (Figure~\ref{fig:workflow}). Based on the input, \textit{\textit{Colin}} generates a textual story outline for parental preview, where parents can directly edit story details through text input. After approval, \textit{\textit{Colin}} generates four scene illustrations according to the storyline for review. \textit{\textit{Colin}} then enters the \textbf{Co-creating Story} module, where the LLM-driven chatbot guides children through two rounds of interactive storytelling. During this process, the GPT-4–powered chatbot expands the storyline in a personalized and engaging way tailored to children's responses (see Section~\ref{sec:prompt_strategy} for prompt design details). \textit{\textit{Colin}} next proceeds to the \textbf{Deeper Understanding} and \textbf{Transfer to Build New Story} modules. Through guided questions, it helps children reflect on and extend the story across three progressive levels—\textbf{Plot}, \textbf{Core Idea}, and \textbf{Transfer}—gradually improving their narrative skills from recalling the script, to understanding key ideas, to constructing new stories. Throughout the process, both newly created and original story elements are stored in the \textbf{Customize Library} module, helping parents review children's creative history and design future story scripts aligned with their interests. 

\subsection{Prompt Design and Refinement}
\label{sec:prompt_strategy}

\begin{figure}[h!]
\centering
\includegraphics[width=\linewidth]{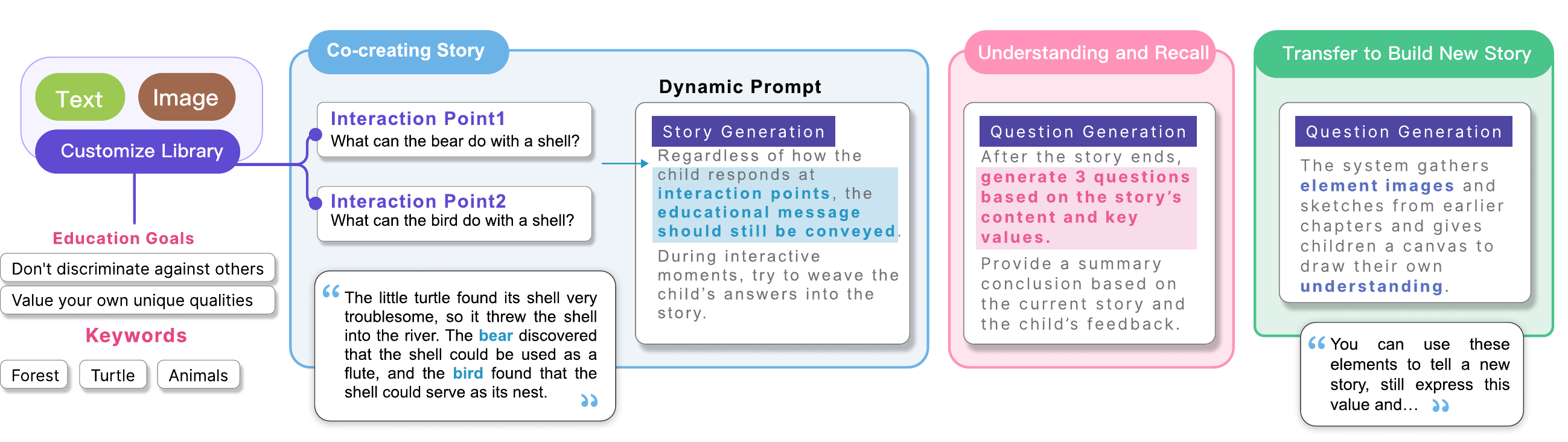}
\caption{Dynamic prompt strategy for AI-assisted storytelling. The system adapts story generation and questioning to children’s responses while maintaining core educational goals.}
\Description{User Interface of \textit{Colin}}
\label{fig:new_prompt_design}
\end{figure}

To prevent the narrative from diverging, ensures the AI maintains narrative coherence while still fostering creativity, a core constraint was embedded in all prompt designs: ``Whenever possible, embed the child's response into the story's progression''. To optimize the prompts for an interactive AI story generator, a pre-experiment was conducted with three children aged 6--8. The goal was to categorize all potential feedback patterns during question-based story interactions. The experiment revealed that children's responses could be classified into three main types: \textbf{precise}, \textbf{vague}, and \textbf{irrelevant}. Based on these feedback patterns, a targeted prompt strategy was designed for each scenario:

\begin{enumerate}
    \item \textbf{Handling Precise Answers (Precise Prompting):} When a child provides a clear, direct answer (e.g., ``Use the shell to make a little drum''), the AI directly incorporates the idea to continue the story.
    \item \textbf{Handling Vague Answers (Elaborative Prompting):} For vague responses (e.g., ``Make something you can hit''), the prompt directs the AI to perform ``clarification and specification.'' The AI first affirms the child's idea and then interprets it within the story's context into a more concrete concept (e.g., specifying ``something you can hit'' as a ``small drum'') before generating the subsequent plot and images. This strategy leverages the AI's imaginative capabilities.
    \item \textbf{Handling Irrelevant or Null Answers (Re-engaging Prompting):} If a child says ``I don't know'' or gives an off-topic response, the prompt triggers a ``restatement and guidance'' strategy. The AI re-engages the child by repeating the story's context and the original question, or by posing a new leading question (e.g., ``What do you think the little bear could do with the shell?'').
\end{enumerate}




\subsection{Evaluation of Generated Content}
\subsubsection{Evaluation Method} 
To evaluate the quality of generated story expansions and reflective questions, we conducted a blind comparison experiment inspired by the gdpval framework~\cite{patwardhan2025gdpvalevaluatingaimodel}. We designed a set of identical story outlines and educational points, which were distributed to both human participants and GPT-4. By comparing the content generated by both groups, we evaluated GPT-4's performance in co-creating story plots, formulating open-ended questions about core ideas, expanding storylines, and providing detailed explanations of central concepts. We recruited 20 participants: 15 with professional experience in early childhood education and 5 with parenting experience with primary school children aged 6-12. During a brief pilot study, we observed children's reactions at key story nodes and randomly selected some of their feedback for inclusion in the main questionnaire. The questionnaire presented the designed questions along with three randomly selected instances of children's feedback, requiring participants to expand the plot and explain their reasoning based on these inputs. The design for human participants mirrored the prompts given to GPT-4, ensuring both groups received the same tasks and guidance in the co-creation process. We collected 80 stories from the 15 participants with early childhood education experience. To create a balanced dataset for comparison, we then generated a corresponding set of 80 stories using GPT-4.

\subsubsection{Evaluation Metrics} 
\begin{table*}[t]
  \centering
  \caption{Content and Question Generation Evaluation Metrics}
  \label{tab:combined_metrics_vertical_title}
  { 
    \footnotesize 
    \renewcommand{\arraystretch}{1.1} 
    \begin{tabularx}{\textwidth}{@{} p{2.5cm} l X @{}}
      \toprule[1pt]
      \bfseries Metric Category & \bfseries Evaluation Dimension & \bfseries Detailed Explanation of Metrics \\
      \midrule
      \multirow{5}{=}{\bfseries Content Generation Metrics} 
      & Preference & Whose story/principle do you like better? \\
      & Understandability~~\cite{zhang2024mathemyths} & Who explains the key ideas more clearly and concretely for children? \\
      & Logical Rationality & Whose story is more logical, with plot develops more rationally? \\
      & Attraction & Whose story sounds more creative, contains more twists and turns and more vivid details? \\
      & Story Relevancy & Which explanations are more in line with the story's core educational goals? \\
      \midrule 
      \multirow{3}{=}{\bfseries Question Generation Metrics} 
      & Story Relevancy & The questions generated relate to the story plot \\
      & Inspiration & The generated question stimulates children's thinking and provokes them to articulate their thoughts \\
      & Inclusiveness & The questions generated are open-ended, allowing children to answer freely \\
      \bottomrule[1pt]
    \end{tabularx}
  } 
\end{table*}
\vspace*{0.5em}
\begin{figure}
    \centering
    \includegraphics[width=.9\linewidth]{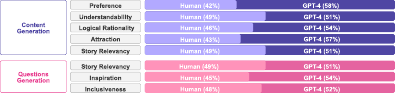}
    \caption{Comparison of human and GPT-4 performance in content and question generation. GPT-4 outperforms humans across most dimensions, including logical rationality, attraction, and inspiration.}
    \label{fig:Evaluation Results}
\end{figure} 

Table \ref{tab:combined_metrics_vertical_title} provides a detailed explanation of these metrics for assessing the story generation and question generation. The metrics are based on a widely accepted approach to evaluating AI-generated conversations presented in the work of Li and Ji et al.~\cite{shi2023large,lovato2019young}. In this method, evaluators were shown stories where one was generated by GPT-4 and another one was generated by a human. They were asked to choose from each pair based on the evaluation dimension. We asked three evaluators to choose the better one from each pair of generated text based on these metrics given in the table~\ref{tab:combined_metrics_vertical_title}. All assessments were conducted by three university student research assistants with more than one year of experience in the education of children.

\subsubsection{Results} 
As shown in Figure~\ref{fig:Evaluation Results}, GPT-4–generated content performed comparably to human-authored content in terms of intelligibility and exceeded it in logical coherence, engagement, and overall preference. Interestingly, human-generated questions received lower scores for story relevance, which may be attributed to participants interpreting the questions in relation to broader life experiences rather than the specific story context.


\section{User Study}


\subsection{Participants}
We recruited 20 pairs of participants from child training institutions and kindergartens through offline recruitment. Each pair consisted of a parent and a child. The child participants included 11 females, 9 males, aged 5 to 13 years old (M=7.97, SD=2.81). This experiment was approved by the Institutional Review Board (IRB) of the affiliated university. Each participant received a 20CNY (about 3 USD) compensation for the 60-minute session.

\subsection{Procedure}
Children participants first answered background questions about their storytelling habits, including: (1) age and learning stage, (2) frequency of story listening (e.g., once per week), (3) ability to retell stories, and (4) tendency to add personal ideas. Before the formal session, a research assistant introduced the interface, demonstrating the correct use of the voice and canvas buttons and explaining the four icons representing different story stages. After confirming understanding through brief practice questions, participants proceeded to the formal experiment, which consisted of three phases: \textbf{Pre-test}, \textbf{During Task}, and \textbf{Post-test}.

\textbf{Pre-test.}  
This phase assessed children’s baseline understanding of the story content and related knowledge before interacting with \textit{Colin}. Each participant read a short story within 10 minutes and answered comprehension questions about its core ideas. They were then asked to identify key thematic words summarizing the story’s main message and to orally create a new story with a different plot but a similar theme. Measures of speech fluency and content association were recorded.

\textbf{During Task.}  
Using an iPad, participants interacted with \textit{Colin} across three structured stages, completing two voice-based and two drawing-based co-creation tasks, two core-value questions, and one transfer task that required creating a new story using prior visual elements. A research assistant was present to provide minimal procedural support when necessary but refrained from offering direct guidance on story content.

\textbf{Post-test.}  
Following the interaction, participants completed a 5-point System Usability Scale (SUS) questionnaire and a 20-minute semi-structured interview to reflect on their experience and perceptions of \textit{Colin}.


\subsection{Evaluation Metrics}
\subsubsection{Narrative Skill Evaluation}
Participants evaluated the storytelling system using a 5-point System Usability Scale (SUS) adapted for children (Cronbach’s $\alpha$ = 0.78). All items were rewritten in age-appropriate, colloquial language. To improve visual comprehension, the traditional 5-point Likert scale was replaced with a pictorial version ranging from crying to smiling faces, representing ratings from ``very unhappy'' to ``very happy.'' The full scale is provided in Appendix~B. Given preschoolers’ limited literacy, the research assistant read each item aloud, displayed the visual scale on screen, and explained its meaning before asking the child to select a response. Following the scale assessment, participants completed a 20-minute semi-structured interview on their experiences with \textit{Colin}. Example questions included: ``Did \textit{Colin} help you think about relationships between story characters?'', ``Would you like to use \textit{Colin} again for storytelling practice?'', and ``Was any part of the system difficult to understand?''

\begin{figure}[h!]
    \centering
    \includegraphics[width=0.8\linewidth]{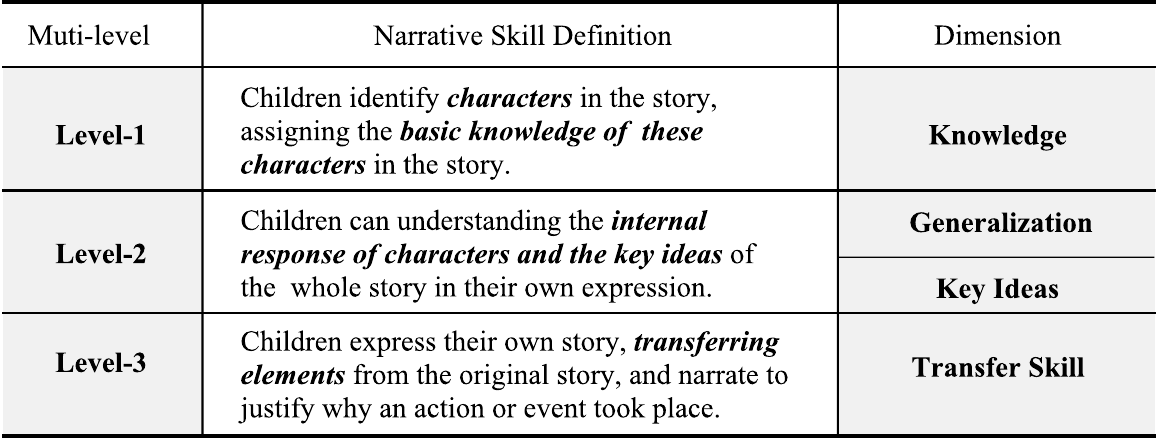}
    \caption{Three-level framework for evaluating children's narrative skills, including knowledge, key idea understanding, and transfer ability.}
    \label{fig:evaluation}
\end{figure}

To assess the effect of \textit{Colin} on children’s narrative understanding and skill development, we drew upon the Language Arts standards from the U.S. Common Core~\cite{englishlanguagestandard} and narrative assessment frameworks from prior research~\cite{petersen2008emerging}. As shown in Figure~\ref{fig:evaluation}, we evaluated three levels of narrative ability—\textbf{Knowledge}, \textbf{Key Idea}, and \textbf{Transfer Skill}—through pre- and post-test comparisons. To reduce repetition bias, post-test questions were rephrased slightly from the pre-test versions. A full list of example questions is included in Appendix~\ref{tab:questions}.

The \textbf{Knowledge} dimension assessed factual comprehension of story elements (e.g., ``What are the characteristics of a turtle’s shell?''). Each story contained 3–5 such items. The \textbf{Key Idea} dimension examined understanding of the story’s central theme or moral. The \textbf{Internal Response} dimension evaluated children’s ability to analyze character motivations and propose alternative plot developments. All \textbf{Narrator Evaluation} items were open-ended, allowing children to express their reasoning freely.

\subsubsection{Engagement and Enjoyment}
We measured children's engagement in two ways, behavioral observations and quantitative analysis. Behavioral observation was used because child subjects sometimes have difficulty expressing their thoughts as clearly as adult subjects. Therefore, behavioral experiments can also give real feedback on children's participation in the \textit{Colin}.
\begin{table}[htbp]
\centering
\caption{Engagement and Enjoyment Evaluation and Semi-structured Interview Guideline}
\renewcommand{\arraystretch}{1} 
\begin{tabularx}{\textwidth}{>{\small}lX} 
\toprule
\textbf{Engagement assessment} & \textbf{Guidance} \\
\midrule
\multirow{4}{*}{\parbox{3cm}{\centering\textbf{Behavioral Observations}}} & Whether the expressions and movements are happy during the entire process of participating in the experiment. \\
& Whether the tone of voice is positive when interacting with \textit{Colin}. \\
& Whether the experiment requires repeated use of certain functions of \textit{Colin}. \\
& Whether children are willing to discuss \textit{Colin} with the experimenter after the experiment. \\
\midrule
\multirow{4}{*}{\parbox{3cm}{\centering\textbf{Quantitative Analysis}}} & Whether the child is happy when telling a story with \textit{Colin}. \\
& Whether the child feels comfortable telling a story with \textit{Colin}. \\
& Whether the child wants a reading companion like \textit{Colin}. \\
& Whether the child wants to tell another story with \textit{Colin}. \\
\bottomrule
\end{tabularx}
\end{table}
To measure children's enjoyment in interacting with artificial intelligence or humans, we adapted a questionnaire consisting of 4 items based on the prior work of Waytz~~\cite{waytz2014mind}.
The experimenters verbally presented questions to the participants and guided them to indicate their level of agreement using a graphical Likert scale. The 5-point scale was visualized as five circles, each depicting a different facial expression: "Strongly Agree", "Agree", " Neither Disagree Nor Agree", "Disagree", and "Strongly Disagree". The circles varied in size, with the smallest representing the "Strongly Disagree" (crying face) and the largest representing the "Strongly Agree" (laughing face). Researchers pointed to each circle when describing the corresponding option , which helped children better understand what each score represented.\cite{hall2016five} The Cronbach's alpha for internal consistency was 0.77, indicating acceptable reliability.

\subsection{Data Collection and Analysis}

The pre-test and post-test interviews were audio-recorded and transcribed for analysis. We employed reflexive thematic analysis~\cite{Braun01012006} to examine the qualitative data. The process consisted of four steps. First, after each user study, researchers made detailed field notes on participants’ behaviors and responses during the interaction. Two researchers independently performed open coding on these notes, discussed discrepancies, and inductively refined a set of initial codes. Next, one researcher applied these codes to all interview transcripts to ensure comprehensive coverage. Both researchers then reviewed and iteratively refined the codebook through discussion until consensus was achieved. Subsequently, the codes were clustered into categories, and overarching themes were derived in relation to the research questions. All qualitative analysis was conducted in Chinese, and representative quotations were later translated into English for reporting.

\begin{figure}[!htbp]
    \centering
    \includegraphics[width=1\linewidth]{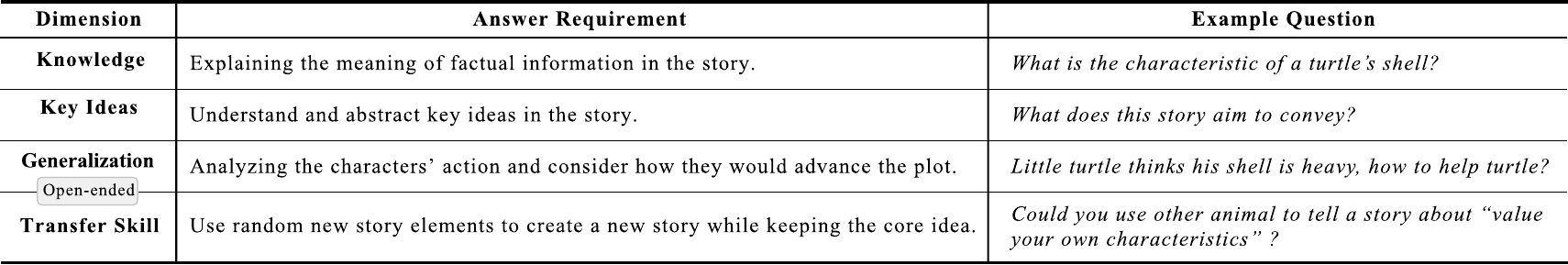}
    \caption{Four dimensions of reading comprehension: Knowledge, Key Ideas, Generalization, and Transfer Skill with example questions.}
    \label{fig:enter-label}
\end{figure}

For quantitative evaluation, three research assistants with experience in early childhood education independently rated children’s performance according to the indicators defined in Figure~\ref{fig:enter-label}. The \textbf{Knowledge} and \textbf{Key Idea} dimensions were scored dichotomously (0 = incorrect, 1 = correct). The \textbf{Internal Response} and \textbf{Narrator Evaluation} dimensions used a 3-point rubric: 2 for successfully applying a key idea to a new story, 1 for retelling the original story in their own words, and 0 for failing to complete the task. Inter-rater discussions were conducted to ensure scoring consistency.

\subsection{Findings}

\subsubsection{Learning Outcome}
Since the data did not meet the assumption of normality, pre-test and post-test comparisons were analyzed using the paired-sample Wilcoxon signed-rank test. In the \textbf{Key Ideas} dimension, all items and total scores yielded $p < 0.001$, indicating significant improvement. This suggests that co-creating stories with \textit{Colin} effectively enhanced children’s understanding of the core messages embedded in the stories. Similarly, the \textbf{Internal Response} dimension showed significant pre–post differences across all items and overall scores, suggesting that \textit{Colin} supported children in not only understanding but also abstracting and generalizing narrative concepts.

In the \textbf{Narrator Evaluation} dimension, eight participants successfully created new stories that conveyed the same core ideas as the original narratives shortly after the session. Nine participants were able to retell the original story to express the learned moral, while only three of the youngest participants (aged 5–6) were unable to construct their own stories. These findings suggest that most children could transfer narrative understanding to novel contexts.

\begin{figure}[t]
\centering
\includegraphics[width=0.7\textwidth]{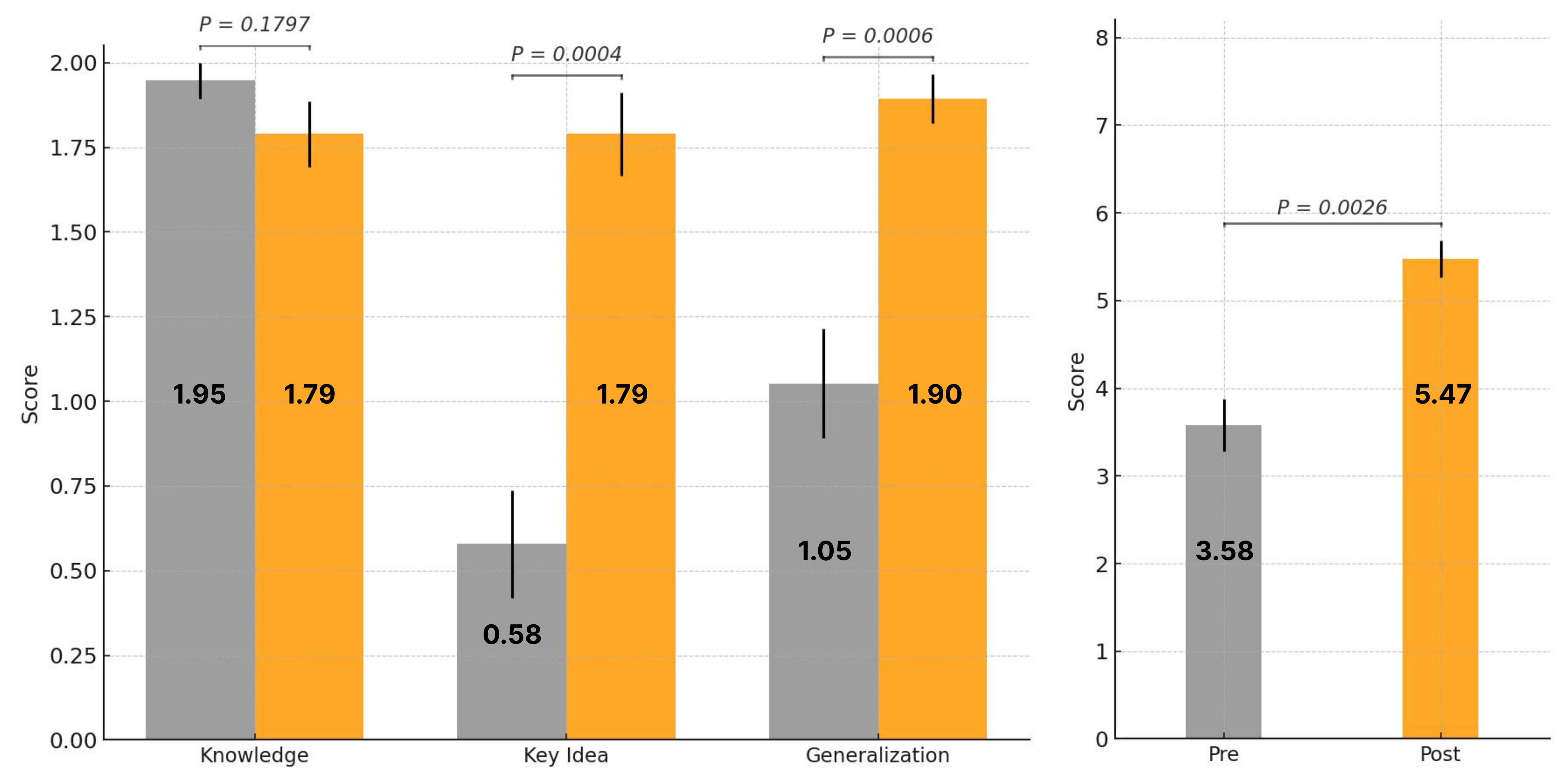}
\caption{Comparison of mean scores (Mean ± SEM) between pre-test and post-test. Gray bars represent pre-test scores, and orange bars represent post-test scores. The black numbers inside each bar indicate the mean scores for that condition. The left panel shows the average scores for Knowledge, Key Idea, and Generalization (maximum score = 2), while the right panel shows the Total score (maximum score = 8). Error bars indicate the standard error of the mean (SEM).}
\label{fig:pre_post}
\end{figure}

As shown in Figure~\ref{fig:pre_post}, the median scores across all dimensions increased, with particularly notable gains in \textbf{Key Ideas} and \textbf{Internal Response}. The overall composite score, calculated from the four learning indicators, also demonstrated an upward shift in the median, indicating that \textit{Colin} effectively fostered multi-level narrative skill development.

Figure~\ref{fig:pre_post} summarizes the statistical outcomes. Significant improvements were observed in three of the four dimensions—\textbf{Key Ideas}($p < 0.001$), \textbf{Generalization}($p < 0.001$), and \textbf{Transfer Skill}—($p < 0.001$). The \textbf{Knowledge} dimension, however, did not show a significant change ($p > 0.18$), likely because the story materials featured familiar elements (e.g., common animals, foods, and plants) that limited measurable learning effects in this category.


\begin{figure}
    \centering
    \includegraphics[width=1\linewidth]{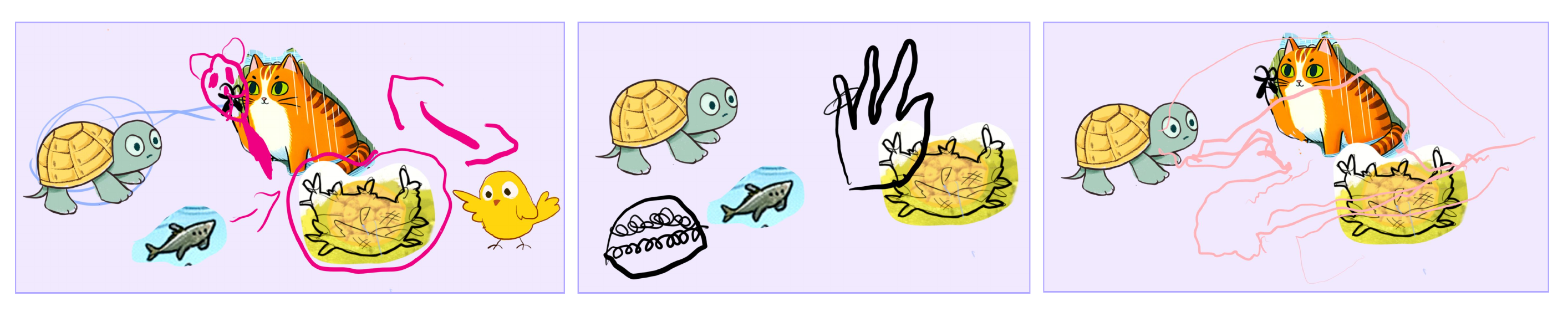}
    \caption{Children’s performance in the transfer stage, freely using lines to build logical connections between story elements (including both elements from the learned story and new ones co-created by the children and completed by AI).}
    \label{fig:placeholder}
\end{figure}

\subsubsection{Learning Experiences to \textit{Colin}}

We conducted semi-structured interviews with all participants, including parents and children to observe their attitudes towards the \textit{Colin}.

\textbf{Quantitative analysis} Among the quantitative indicators, all subjects were able to give answers. Among all the recorded answers, 30.6\% were meta-comments reflecting the thinking process, and 29.22\% contained substantive statements with explanatory significance. The remaining answers that were either irrelevant to the question, or uncertain, or refused to complete the interactive task accounted for 9.8\%, and the short answers of less than five words accounted for 39.2\%, the highest proportion, possibly because many children answered the plot questions directly with a single word. Therefore, in terms of the combination of quantitative indicators and behavioral observation, most of the subjects actively participated in the process of co-creating stories and interacting with \textit{Colin}.

\textbf{Behavioral observation of Engagement} The observations include whether participants showed happiness (16/20), interacted with \textit{Colin} in a positive tone (16/20), repeatedly used certain \textit{Colin} features (12/20), and whether they were interested in continuing the discussion about \textit{Colin} after the experiment (8/20). 

\begin{figure}[!htbp]
\centering
\includegraphics[width=\textwidth]{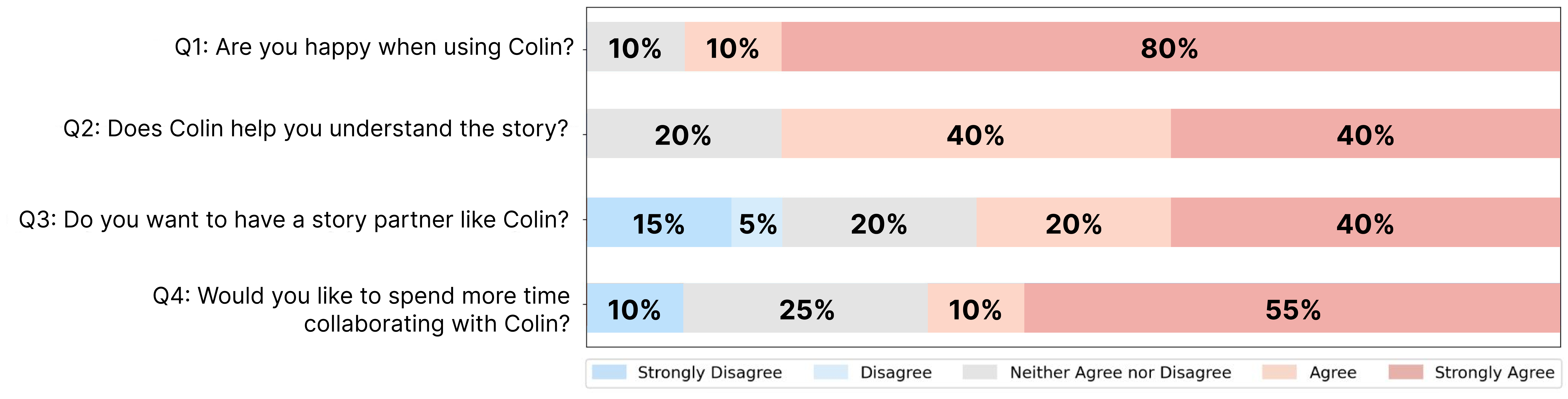}
\caption{Children's questionnaire responses to four evaluation questions about \textit{Colin}. Red shades represent agreement (light red = Agree; dark red = Strongly Agree), blue indicates disagreement (Strongly Disagree), and gray represents a neutral response (Neither Agree nor Disagree).}
\label{fig:happiness}
\end{figure}

\textbf{Evaluation of Enjoyment} The figure\ref{fig:happiness} shows the children's responses to four questions (Q1 to Q4) about the pleasantness of the system. Each bar represents the percentage of responses ranging from ``strongly disagree" to ``strongly agree." For Q1, a majority (84\%) strongly agreed. Feedback from Q2 was more balanced, with 42\% agreeing. Q3 shows that a majority of 21\% neither agree nor disagree (gray) and 42\% agree. 57\% of Q4 respondents agreed, with the rest of the responses spread across other categories. Red, orange, gray, light blue, and dark blue represent different levels of disagreement and agreement, respectively.

Children demonstrated an overall positive perception of learning story content through collaborative storytelling with \textit{Colin}. Engagement and satisfaction levels were consistently high. Using a five-point Likert scale (1–5) to assess willingness and satisfaction, 16 out of 20 participants rated all four dimensions with the highest score of 5, indicating strong enjoyment and motivation during interaction with the AI story assistant.
\section{Discussion}
Our discussion is organized around three core research questions: 

How Large Language Models (LLMs) can be effectively and appropriately used to support facilitator-involved scaffolding and co-creation in children's narrative development \textbf{(RQ1)}. 

How the system's adaptive interaction mechanisms influence children's learning outcomes and experiences \textbf{(RQ2)}. 

How multi-stakeholders—children and parents—perceive this LLM-empowered storytelling system \textbf{(RQ3)}.

\subsection{RQ1: How can LLMs be effectively and appropriately used to support facilitator-involved scaffolding and co-creation in children's narrative development?}
To effectively and appropriately use LLMs to support children's narrative development, our research finds that LLMs provide scaffolding in two primary ways: first, by expanding children's imaginative boundaries and overcoming the limitations of human facilitators; and second, by generating high-quality interactive content that offers concrete support for educational practices. However, using them "appropriately" also requires acknowledging and addressing their inherent challenges.

\textbf{Expanding Imaginative Possibilities Beyond Human Limitations}
A key advantage of LLMs identified in our study is their ability to unboundedly accommodate, embed, and expand upon children's imagination—a feat often difficult for human facilitators like parents and teachers. One parent noted that his own thinking might limit his child's imagination, whereas an LLM would perform better in this regard:
\begin{itemize}
    \item[] \textit{``Later, when she grew up, my daughter would read stories by herself and rarely read stories with us, because I felt that the child would think that we might hinder some of her understanding or some of her ideas. Because now when she grows up, she has her own thoughts. Just like when she encounters a fork in the story, she can develop from direction A or direction B. Her ideas may conflict with ours, so AI can accompany her to explore on her own, and the AI can completely extend the story according to her ideas."}
\end{itemize}
This approach, where children create directly without knowing the full plot, helps them concentrate on plot development and fosters divergent thinking and imagination. Compared to human guides, the system's advantage lies in its ability to continuously embed children's ideas into the narrative without requiring a high level of background knowledge from the facilitator. In a classroom, a teacher cannot generate a unique story for every child, but an AI can serve as an excellent creative assistant.

Both preschool teachers mentioned that adult thinking, shaped by social conventions, can become solidified, leading to strongly guided storytelling that may constrain a child's imagination. As one preschool teacher noted:
\begin{itemize}
    \item[] \textit{``When the mother tells a story, she may only guide one route, and the children will only know this one route after listening to the story. AI can help children think about the story from many different angles."}
\end{itemize}
For parents, the process of repeatedly telling the same story while replacing elements can be tedious. \textit{Colin} can automate the process of exploring story possibilities for children.

\textbf{Providing High-Quality Scaffolding for Educational Practices}
Beyond expanding imagination, LLMs can provide high-quality scaffolding in educational interactions. In our model evaluation, we designed a questionnaire where 20 participants (15 with early childhood education experience) were asked to devise interactive questions and plot expansions based on a story outline. Participants commonly found the task "challenging to answer," with a median completion time of 1638 seconds. This highlights the real difficulties facilitators face in co-creating stories with children.

In contrast, GPT-4 excelled at generating interactive content. Its questions were more inspiring and inclusive, better at promoting story development and gathering meaningful feedback. Human participants sometimes created questions more related to daily issues or educational goals, often resulting in yes/no questions that limited children's response opportunities. GPT-4's questions, however, were more closely aligned with the plot and encouraged comprehensive answers.

GPT-4 also demonstrated superior performance in attraction and explanatory clarity when expanding plots, likely due to its ability to create tight logical connections based on specific instructions. While human-generated content was comparable in understandability and story relevance, evaluators noted that AI-generated stories sometimes lacked internal logical consistency despite their syntactically sound structure, aligning with existing research~\cite{shi2023large,lovato2019young}.

\textbf{Challenges and Appropriate Use of LLMs}
To use LLMs ``appropriately," one must also address their inherent challenges, which form an important part of our discussion.

\textit{Unpredictability of LLM Output.} Our design process revealed that LLMs exhibit a high degree of variability: minor changes in prompts often led to substantially different story continuations. This unpredictability highlights the ``black-box'' nature of prompt interpretation. While extensive testing during development improved stability, achieving reliable control over story quality remains an open challenge. This variability, however, also suggests opportunities for harnessing LLM creativity, as diverse outputs can enrich children's imaginative exploration when appropriately curated.

\textit{Fine-grained Control and Instruction Following.} Although LLMs handle general narrative constraints well, they struggle with precise numerical or structural requirements (e.g., producing openings between 100–200 words). Such limitations reflect broader gaps in LLMs' reasoning with quantitative constraints. Building validation layers~\cite{he2023solving} or feedback loops~\cite{lee2022interactive, lee2023towards, lee2023dapie, lee2024open} into storytelling systems may help ensure outputs align with intended pedagogical and design goals. Moreover, while deviation from prompts occasionally disrupted plot coherence, in practice, this created unexpected narrative turns that some children found engaging. Balancing reliability and creative surprise thus represents an important design space for future research.

\textit{Hallucinations and Educational Implications.} We also observed instances of LLM hallucinations—plausible yet logically flawed content, such as describing how turtle shells could be used to make herbal jelly. In the context of Colin's value-oriented education, these errors did not undermine learning outcomes. However, in domains requiring factual accuracy, such hallucinations could be problematic. Emerging approaches, such as post-hoc filtering of AI outputs~\cite{dziri2021neural} or grounding generation in external knowledge~\cite{ren2023investigating}, may mitigate these risks. At the same time, such phenomena provide opportunities to engage children in critical thinking, for instance, by prompting them to identify and discuss implausible elements in AI-generated stories.

\subsection{RQ2: How can adaptive interaction mechanisms influence children's learning outcomes and learning experiences?}
The system's adaptive interaction mechanisms positively influenced children's learning experiences and outcomes by enhancing engagement, ownership, and immersion. However, our observations also reveal that achieving broader effectiveness requires a deeper level of personalization.

\textbf{Enhanced Learning Experience: Engagement, Ownership, and Immersion}
Children showed great enthusiasm for the co-creation model. Twelve participants expressed a desire to use \textit{Colin} repeatedly. The most frequently requested functionalities were the drawing feature during plot creation, regenerating expanded text, and the drawing feature in the Narrator Evaluation section. The drawing feature was particularly popular, likely because the concrete act of drawing with a stylus enhances children's sense of participation and creativity with a low barrier to entry.

Deep participation in the story creation process fostered a sense of ownership, making children feel that it was "my story." As one parent analogized:
\begin{itemize}
    \item[] \textit{``I think it's good to do it in the same way that when we play RPGS (Role-Playing games), we have some plot options. When the user chooses this plot, the story goes in this direction. It makes the player feel very involved and immersed."}
\end{itemize}
This sense of ownership made children care more about the stories they helped create. They also enjoyed the repeated, changing feedback; rather than a one-way output, the AI assistant fine-tuned its extensions based on the child's current input, enhancing the adaptive experience.

\textbf{Improved Learning Outcomes: Deeper Understanding and Focused Attention}
The positive learning experience directly contributed to better learning outcomes. Participants agreed that co-creating the story was conducive to understanding its key ideas. One parent noted that as \textit{"children guide the development of the story... it is a process of constantly understanding the story background and thinking about the direction of the story plot,"} believing this deep participation led to deeper understanding.

Another parent observed that the interactive mechanisms significantly improved their child's focus:
\begin{itemize}
    \item[] \textit{``When listening to the story background, he may still listen to the recording as before. When he encounters the first problem, he will think back to what happened before... from his first question to his second question, he will listen or watch more carefully than he did before. Because he knew there would be a second question, and he wanted to shape the story... so he will listen more carefully."}
\end{itemize}
This heightened attention, driven by the desire to control the narrative, helped children develop a deeper impression of the knowledge points within the story. The addition of visual creation also effectively captured children's attention in a way that traditional one-way storybooks or recordings often do not.

\textbf{The Need for Deeper Personalization: Adapting to Individual Differences}
While our findings indicate that the system's interactive mechanisms generally enhance engagement and learning, we also observed that their effectiveness is not uniform across all children. \textit{A key insight from our study is that children's individual characteristics—such as personality (introversion vs. extroversion), existing narrative skills, and family background—significantly mediate their learning outcomes.}

For instance, two six-year-old girls outperformed many older children, particularly in the "transfer" stage. They reacted faster, created stories more relevant to the core values, and used new material rather than retelling the original story. The parents of both children are university professors, suggesting that a supportive home environment and pre-existing articulation skills may allow some children to better leverage the creative freedom offered by the system.

This observation implies that a one-size-fits-all adaptive model may be insufficient. As education experts mentioned in interviews, children at different developmental stages (e.g., 3-6, 6-8, and 8+) have vastly different narrative and logical abilities. Therefore, future adaptive systems should not only respond to in-the-moment user input but also provide differentiated scaffolding based on a child's broader profile. For children who are more introverted or slower to respond, the system might need to offer more structured, detailed support, especially during challenging creative tasks. This points to the necessity of designing systems that can track a child's progress over time and adjust interaction difficulty and support levels accordingly, creating a truly personalized learning trajectory.

\subsection{RQ3: How do multi-stakeholders (children and parents) perceive an LLM-empowered storytelling system in supporting children's narrative development?}
As the primary stakeholders, both children and parents perceived unique value in the system from their respective viewpoints.

\textbf{Children's Perspective: An indefatigable and Personalized Companion}
Children expressed high satisfaction with the combination of voice and visual modalities. Two participants, aged 5-6, accurately articulated the value of the AI story assistant without guidance from experimenters:
\begin{itemize}
    \item[] \textit{``It can save my parents' energy because it is a machine and can tell me new stories unlimited times."}
    \item[] \textit{``My dad is very busy with work, he only tells me once, but AI can tell me many times."}
\end{itemize}
From the children's perspective, the system is an indefatigable and endlessly novel storytelling companion that can tell stories unlimited times, fulfilling their dual needs for repetition and novelty.

\textbf{Parents' Perspective: A Powerful Tool for Scaffolding and Assessment}
Parents, in contrast, viewed the system from a more pedagogical standpoint. They saw it as a tool that not only sparked imagination but also effectively exercised core narrative skills. A parent of a 9-year-old child recognized the assessment significance of the final stage:
\begin{itemize}
    \item[] \textit{``I think this link is quite meaningful. Children can organize stories in their own language, exercise their expression skills, use their imagination, and test her ability to weave stories around key ideas is very important."}
\end{itemize}
Overall, parents perceived the system as a powerful auxiliary tool that could foster creativity in ways that humans cannot, while also helping to develop and assess their children's expressive and logical abilities, thus effectively supporting their educational practices at home.

\section{Limitations and Future Work}

\subsection{Scalability and Longitudinal Effects}
The participants in our study engaged in only a single session of AI-assisted collaborative storytelling. Such short-term interactions may not fully capture the changes that might occur as children interact with the storytelling system over an extended period. With time, children may become more accustomed to co-creating with the system, potentially requiring more adaptive and engaging experiences to maintain their attention. Future research should conduct longitudinal studies to explore the long-term effects of interaction with the system, observing potential shifts in interaction patterns, learning habits, and thinking processes.

\subsection{Generality Across Different Age Groups and Learning Objectives}
The learning objectives for children typically vary across different age ranges such as 3-6, 6-8, and 8+ years. Younger children (3-6) often focus on specific daily habits, while those aged 6-8 begin to understand character motivations and general values. Older children (8+) engage with more complex background knowledge and multi-dimensional character analysis. The current system does not sufficiently differentiate its generated content and interactive questions based on these age-specific learning goals. Future iterations should refine the content generation to align more closely with the developmental needs of distinct age groups and generate questions that target their respective learning focuses. This could involve creating different story interaction modes with varying levels of difficulty and support.

\section{Conclusions}
In this paper, we introduce \textit{Colin}, a child-story creation system that emphasizes key story ideas through the use of a question-scaffolding feedback framework. The storytelling system adopts an interactive model of ``Question-Feedback-Story Generation'', presenting open-ended questions to elicit narrative contributions from children. Based on children's ideas and the story framework, the system expands details to support further story development. The system evaluation reveals that \textit{Colin} enhances multi-level narrative skill for children in story understanding.

\section*{References}


\begin{appendices}
\section{APPENDIX}
\subsection{Prompt Design}
\subsubsection{Outline Generate Prompt}
\label{Outline Generate Prompt}
\ 
\newline
You will receive a story. Use the provided keywords to simplify the story 

Keywords: \{keyword\}

Story: \{story\_text\}

Target: \{target\}

\noindent Where story is the story input, target is the educational goal of the story, and keyword is the character element that needs to appear in the story.

\noindent You need to generate a 3-4 sentences story outline, called a summary, in which only one character appears in each paragraph and describes its actions.

\subsubsection{Story Split Prompt}
\label{Story Split Prompt}
\ 
\newline
\noindent You will receive a story. Please divide the story into four parts:  the beginning, key plot point 1, key plot point 2, and the ending.

\noindent For each part, describe the actions of the characters in the scene.

\noindent The story is: \{story\_text\}

\noindent The four parts are identified with part1, part2, part3, and part4 as keys.
\noindent Each part only needs 1-2 sentences, each part only needs to describe the plot, do not add additional description.

\noindent Here is a complete breakdown of the process

\noindent Outline: The little tortoise found his shell troublesome, so he threw it into the river. The little bear found that the shell could play the flute. The little bird found that the shell could be used as its nest. They all thanked the little turtle, the little turtle felt very happy ", this story educates children "do not discriminate against others", "cherish their own characteristics"

\noindent Return Json format:

\noindent \{

"part1": "Once upon a time, there was a little turtle who always felt that his shell was heavy and hard, which was very troublesome. One day, he decided to take off his shell and throw it into a nearby river",

"part2": "Soon after, Little Bear found the shell while playing by the river. Little Bear ponders the function of this turtle shell",

"part3":"Little Bear picks up the shell, finds it round and has many holes, and suddenly has a good idea. It took the turtle shell as a special instrument and played a melodious sound, just like playing the flute. My friends in the forest were attracted to the beautiful music.",

"part4":"The bird uses the turtle shell as a natural skateboard, gliding between the branches, attracting the envious eyes of other birds, everyone wants to try this special skateboard. The little turtle felt very warm and proud when he heard that his shell had brought so much happiness to his friends."

\noindent \}

\noindent Note: Only one character can appear in each part and do some actions. Don't have multiple character descriptions and interactions in one part

\subsubsection{Background Extract Prompt}
\label{Background Extract Prompt}
\ 
\newline
\noindent You will receive a story's part. Identify all characters and specify what they are.

\noindent For example, If a character is Wukong, who is a monkey, refer to him as Wukong the monkey.

\noindent Assign each character characteristics that describe what they look like, considering the story's context.

\noindent For humans, include descriptions of their clothing. Write a single sentence that depicts an image for this page in the following format:

[subject], [doing action], [adjective], [background subject], [scenery]

\noindent For example: Little monkey with brown fur, curious eyes, entering the cornfield, tall corn stalks, golden field, looking at the big corn, plentiful corn stalks, sunlit field.
The story is: \{story\_text\}

\subsubsection{Character Extract Prompt}
\label{Character Extract Prompt}
\ 
\newline
\noindent You will receive a story. Find all of the characters within this story. Don't just say the name, include what they actually are, without including adjectives. If the character is wukong, who is a monkey, output it as wukong the monkey. 

\noindent Output them in the following format: Character 1 | Character 2 | Character 3 | . 

\noindent The story is : \{story\_text\}

\subsubsection{Story Expand And Interaction Question Generate Prompt}
\label{Story Expand And Interaction Question Generate Prompt}
\ 
\newline
\noindent I need to tell stories to children aged 3-8. Please expand and tell the story. \{story\_split\_text\}

\noindent In the story, you design interaction point 1 and interaction point 2, so that interaction point 1 appears after part2, and interaction point 2 appears after part3.

\noindent Each expansion is less than 250 words and conveys an educational message regardless of the feedback from interaction points 1 and 2.

\noindent In the interaction point, try to embed any of the child's answers in the development of the story. If the answer does not fit into the story at all, or has nothing to do with the question, repeat the question generated at the interaction point.

\noindent If the child has strong resistance to the story telling activity itself, use a patient tone of persuasion and repeat the questions generated by the interaction points.

\noindent Now when we get to interaction point 1 in Parts 1 and 2, you need to stop and wait for my feedback. Your reply should be returned in Json format \{"key":value,... ,"key":value\} :

1. The Json tag name corresponding to the story text is story

2. The text of the question raised in interaction point 1 corresponds to the Json label interact

3. Avoid suggestive text in parentheses

4. Animals, people do not appear names, directly use nouns

\subsubsection{Interaction Feedback Prompt}
\label{Interaction Feedback Prompt}
\ 
\newline
\noindent For interaction point, the answer is {message}, please generate feedback, and pay attention to the following points when generating feedback:

1, timely to the child affirmation and praise

2. Help your child fill in the details of this answer

\noindent After feedback is generated, talk about next part until interaction point stops, waiting for the child's feedback. Your reply should be returned as a pure json string in the format \{"key":value,... ,"key":value\} :

1. The story text and feedback corresponding to the story is named Story

2. The text of the question raised in point is labeled interact

3. Avoid suggestive text in parentheses

4. Animals, people do not appear names, directly use nouns

\subsubsection{Story End And Recall Question Generate Prompt}
\label{Story End And Recall Question Generate Prompt}
\ 
\newline
\noindent After generating feedback, proceed directly to the ending of part4
After the ending, please generate three questions according to the material of the story and the truth conveyed. After each question is asked, you need to stop and wait for my feedback, and give feedback according to the answer content, which is controlled within 200 words.

\noindent In the follow-up, the first question given by children will be question 1 (Q1), followed by question 2 (Q2), and question 3 (Q3). After the answer of Q3, the guidance will describe the summary and feedback of all links.

\noindent If the child's answer does not correctly understand the truth conveyed by the story, you need to patiently guide the child to understand the truth and ask again. Make sure your child understands, then ask the next question. Your reply should be returned in Json format

\subsubsection{Recall Feedback Prompt}
\label{Recall Feedback Prompt}
\ 
\newline
\noindent For the current problem, the child responds as {message}, generates feedback and waits for the child's response to the next question (Q1, Q2, Q3), and the return should be returned in json format. The format is \{"key":value,... ,"key":value\}

1. The feedback generated after answering the questions is labeled guidance

2. The leading text for the next question is labeled interact

3. Avoid suggestive text in parentheses

4. Animals, people do not appear names, directly use nouns
Your return contains only json strings and nothing else

\subsubsection{Post-Study}
\begin{table*}
  \caption{List of pre and post test questions}
  \label{tab:questions}
  \resizebox{\linewidth}{!}{
  \begin{tabular}{ccl}
    \toprule[2pt]
    Evaluation dimension &Problem list(Correctly marked as 1 point, incorrectly marked as 0 points)\\
    \midrule
    Knowledge & 1. Is the turtle's shell light or heavy? \\
    & 2. From the following words, please choose the characteristics of the turtle shell: light/soft/heavy.\\
    \midrule  
    Key idea & 3. Why did the little turtle dislike his shell at first and then become happy again?\\
    & 4. If the baby turtle is sad because its shell is heavy, how do you comfort it?\\
    \midrule  
    Internal Response & 5. What does it mean to cherish?\\
    & 6. What is the definition of characteristics?\\
    \toprule[2pt]
    Evaluation dimension & Problem list(This dimension is scored 0 points in the pre-test, and 0, 1, and 2 points in the post-test respectively)\\
    \midrule
    Narrator Evaluation & 7. Use the elements in the picture (turtle, firefly, cat, fish, bird's nest) to tell a story and express "cherish your own characteristics".\\
    & 8. Using the elements in the picture (turtle, firefly, cat, fish, bird's nest), tell a story about "finding the good in your friends".\\
  \bottomrule[2pt]
\end{tabular}
}
\end{table*}

\begin{table*}[t]
  \caption{Interview questions regarding the storytelling teaching method}
  \label{tab:interview-questions}
  \resizebox{\linewidth}{!}{
  \begin{tabular}{ll}
    \toprule[2pt]
    \textbf{Dimension} & \textbf{Questions} \\
    \midrule
    Specific Teaching Methods &
    \begin{minipage}[t]{12cm}
        \begin{itemize}[leftmargin=*, topsep=3pt, itemsep=2pt, parsep=0pt]
            \item What kinds of stories do you prefer to choose? What are the corresponding teaching objectives for each?
            \item Can you provide a specific example of a lesson taught using the storytelling method?
            \item How many lessons does a single story span? What tasks are completed in each lesson?
            \item When using the storytelling method, how do you interact with the children?
        \end{itemize}
    \end{minipage} \\
    \midrule
    Advantages of the Method &
    \begin{minipage}[t]{12cm}
        \begin{itemize}[leftmargin=*, topsep=3pt, itemsep=2pt, parsep=0pt]
            \item What do you perceive as the advantages of the storytelling teaching method?
            \item In what ways do you believe the storytelling method helps enhance narrative skills?
            \item During the teaching process, how can one better interact with children?
            \item How do the frequency and format of storytelling education affect children's learning outcomes?
        \end{itemize}
    \end{minipage} \\
    \midrule
    Existing Problems &
    \begin{minipage}[t]{12cm}
        \begin{itemize}[leftmargin=*, topsep=3pt, itemsep=2pt, parsep=0pt]
            \item What problems do you encounter while preparing a storytelling lesson?
            \item What issues arise during the process of interacting with children to explain the story's content?
            \item How do you guide children to co-create stories? How do you handle diverse feedback from them?
            \item In the process of co-creating stories and teaching, what kind of feedback from children is particularly difficult to guide?
        \end{itemize}
    \end{minipage} \\
    \midrule
    Views on Intelligent Tutoring Systems &
    \begin{minipage}[t]{12cm}
        \begin{itemize}[leftmargin=*, topsep=3pt, itemsep=2pt, parsep=0pt]
            \item Are you familiar with research on AI-assisted education? What are your views on it?
            \item Have you used products like voice assistants or ChatGPT?
            \item Do you believe these products can be integrated with the storytelling teaching method?
        \end{itemize}
    \end{minipage} \\
    \bottomrule[2pt]
\end{tabular}
}
\end{table*}

\begin{table*}[t]
  \caption{The Impact of Storytelling Teaching on Children's Abilities}
  \label{tab:abilities}
  \resizebox{\linewidth}{!}{
  \begin{tabular}{lp{12cm}}
    \toprule[2pt]
    \textbf{Ability} & \textbf{How Storytelling Teaching Enhances Ability} \\
    \midrule
    Comprehension Ability & Understand the reasons for characters' behaviors in the story and comprehend the daily knowledge it conveys. Identify and clarify the logic in the illustrations. \\
    \midrule
    Imagination & Some stories may have open-ended conclusions without clear answers or explanations. This encourages children to think for themselves and imagine the story's subsequent development. This openness can stimulate children's creative thinking and imagination. \\
    \midrule
    Expressive Ability & Children are encouraged to be creative in the development of the story. They can freely propose their own ideas, add new plots and characters, change the story's ending, and fluently express their own ideas about the plot through language or images. \\
    \bottomrule[2pt]
\end{tabular}
}
\end{table*}

\begin{table}[htbp]
\centering
\caption{Comparison of Human and ChatGPT 4 Responses (Prompt: “What can the little bear make with the shell?”)}
\renewcommand\arraystretch{1.2} 
\setlength{\tabcolsep}{6pt} 
\begin{tabularx}{\textwidth}{@{}>{\raggedright\arraybackslash}p{2.8cm} >{\raggedright\arraybackslash}p{4.2cm} X@{}}
\toprule
\textbf{Child’s Input} & \textbf{Human Facilitator} & \textbf{ChatGPT4} \\
\midrule
“Make something that can hit.” &
“By ‘something that can hit,’ do you mean a gun, a shield, or boxing gloves?”
\vspace{0.5em} 
“Alright, then let’s make a pair of boxing gloves with the shell!” &
The little bear carefully observed the shell. Suddenly, inspiration struck—it decided to make a small drum out of it. The little bear found some branches and leaves and cleverly attached them to both sides of the shell, creating a cute hand drum. Every time the sun rose, the little bear would pick up the drum and gently tap it… \\
\midrule
“Make something really cool.” &
“What kind of thing do you mean by ‘cool’?”
\vspace{0.5em} 
“How about making a flute?” (Directly using the answer from the story) &
The little bear rested its head on the shell and thought for a while. Suddenly, its eyes lit up. It cleverly used branches and leaves to make a cool helmet. Every time it wore this helmet, it felt like it had become a superhero of the forest. The little bear ran happily through the woods… \\
\bottomrule
\end{tabularx}
\end{table}

\newcommand{\TopHeavyRule}{\specialrule{1.2pt}{0pt}{3pt}}
\newcommand{\LightRule}{\specialrule{0.4pt}{3pt}{3pt}}
\newcommand{\BottomHeavyRule}{\specialrule{1.2pt}{3pt}{0pt}}

\begin{table}[htbp]
\centering
\caption{The Complete Interaction Flow between colin and the Child}
\renewcommand\arraystretch{1.4} 
\begin{tabularx}{\textwidth}{@{} >{\bfseries\raggedleft\arraybackslash}p{3.5cm} X @{}}
\TopHeavyRule

colin &
\textit{(Opening Scene)} \par
Once upon a time, there was a little turtle... \\

colin &
\textit{(First Plot Question)} \par
"Little friend, what can the little bear make with the shell?" \\

Child &
\textit{(Presses the record button)} \par
"I know bears like to eat honey. It can use the shell as a bowl." \par
\textit{(Presses the record button again to stop recording)} \\

Child &
\textit{(Presses the drawing button)} \par
colin opens the drawing canvas, and the child begins to draw. \\

colin &
\textit{(Drawing Assistance)} \par
Based on the category of the child's input, SketchRNN helps complete the drawing details. \\

Child &
\textit{(Presses the drawing button again to finish drawing)} \\

colin &
Collects the elements drawn on the canvas. \\
\LightRule

colin &
\textit{(First Story Expansion)} \par
"Wow, little friend, you're so clever! Just as you said, the little bear used the turtle's shell as a bowl..." \\

colin &
\textit{(Second Plot Question)} \par
... (Second round of plot questions and creation is omitted) ... \\

colin &
\textit{(Story Conclusion)} \par
"You did a great job, little friend! The little bird used the turtle's shell as a... This experience changed how the little turtle felt about its own shell..." \\
\LightRule

colin &
\textit{(First Moral Question)} \par
"Little friend, why did the little bird thank the little turtle?" \\

Child &
\textit{(Presses the record button)} \par
"The shell is great." \par
\textit{(Presses the record button again to stop recording)} \\

colin &
\textit{(First Moral Guidance)} \par
"The little bird thanked the little turtle because... This tells us that sometimes, even seemingly ordinary features or things can be of great help to others." \\
\LightRule

colin &
\textit{(Conclusion and Re-creation Prompt)} \par
Randomly displays the elements the child just created, along with other story elements: \par
"Little friend, you can use these elements to tell a new story that still expresses the core idea of the one we just made..." \\
\BottomHeavyRule
\end{tabularx}
\end{table}

\end{appendices}

\end{document}